\documentclass[twocolumn,english,superscriptaddress,citeautoscript,showpacs,preprintnumbers,amsmath,amssymb,prb,floatfix,footinbib]{revtex4}
\usepackage[scaled=2]{helvet}
\usepackage[T1]{fontenc}
\usepackage[utf8]{luainputenc}
\usepackage{textcomp}
\usepackage{amsmath}
\usepackage{amssymb}
\usepackage{graphicx}
\usepackage{subscript}
\usepackage{epstopdf}

\makeatletter

\@ifundefined{textcolor}{}
{%
 \definecolor{BLACK}{gray}{0}
 \definecolor{WHITE}{gray}{1}
 \definecolor{RED}{rgb}{1,0,0}
 \definecolor{GREEN}{rgb}{0,1,0}
 \definecolor{BLUE}{rgb}{0,0,1}
 \definecolor{CYAN}{cmyk}{1,0,0,0}
 \definecolor{MAGENTA}{cmyk}{0,1,0,0}
 \definecolor{YELLOW}{cmyk}{0,0,1,0}
}

\usepackage[scaled=2]{helvet}\usepackage{amstext}

\makeatletter
\@ifundefined{textcolor}{}{\definecolor{BLACK}{gray}{0}\definecolor{WHITE}{gray}{1}\definecolor{RED}{rgb}{1,0,0}\definecolor{GREEN}{rgb}{0,1,0}\definecolor{BLUE}{rgb}{0,0,1}\definecolor{CYAN}{cmyk}{1,0,0,0}\definecolor{MAGENTA}{cmyk}{0,1,0,0}\definecolor{YELLOW}{cmyk}{0,0,1,0}}

\usepackage{dcolumn}
\usepackage{bm}
\usepackage{times}
\usepackage{hyperref}
\hypersetup{colorlinks=true,linkcolor=red,citecolor=blue}
\makeatother

\usepackage{babel}

\makeatother

\usepackage{babel}
\begin{document}

\title{Magnetic polarization of Ir in underdoped, non-superconducting Eu(Fe$_{0.94}$Ir$_{0.06}$)$_{2}$As$_{2}$ }

\author{W. T. Jin}

\email{w.jin@fz-juelich.de}

\affiliation{J\"{u}lich Centre for Neutron Science JCNS and Peter Grünberg Institut PGI, JARA-FIT, Forschungszentrum J\"{u}lich GmbH, D-52425 J\"{u}lich, Germany}

\affiliation{J\"{u}lich Centre for Neutron Science JCNS at Heinz Maier-Leibnitz Zentrum (MLZ), Forschungszentrum J\"{u}lich GmbH, Lichtenbergstraße 1, D-85747 Garching, Germany}

\author{Y. Xiao}

\email{y.xiao@fz-juelich.de}

\affiliation{J\"{u}lich Centre for Neutron Science JCNS and Peter Grünberg Institut PGI, JARA-FIT, Forschungszentrum J\"{u}lich GmbH, D-52425 J\"{u}lich, Germany}

\author{Y. Su}

\affiliation{J\"{u}lich Centre for Neutron Science JCNS at Heinz Maier-Leibnitz Zentrum (MLZ), Forschungszentrum J\"{u}lich GmbH, Lichtenbergstraße 1, D-85747 Garching, Germany}

\author{S. Nandi}

\affiliation{J\"{u}lich Centre for Neutron Science JCNS and Peter Grünberg Institut PGI, JARA-FIT, Forschungszentrum J\"{u}lich GmbH, D-52425 J\"{u}lich, Germany}

\affiliation{J\"{u}lich Centre for Neutron Science JCNS at Heinz Maier-Leibnitz Zentrum (MLZ), Forschungszentrum J\"{u}lich GmbH, Lichtenbergstraße 1, D-85747 Garching, Germany}

\affiliation{Department of Physics, Indian Institute of Technology, Kanpur 208016, India}

\author{W. H. Jiao}

\affiliation{School of Science, Zhejiang University of Science and Technology, Hangzhou 310023, China}

\author{G. Nisbet}

\affiliation{Diamond Light Source Ltd., Diamond House, Harwell Science and Innovation Campus Didcot, Oxfordshire OX11 0DE, UK}

\author{S. Demirdis}

\affiliation{J\"{u}lich Centre for Neutron Science JCNS at Heinz Maier-Leibnitz Zentrum (MLZ), Forschungszentrum J\"{u}lich GmbH, Lichtenbergstraße 1, D-85747 Garching, Germany}

\author{G. H. Cao}

\affiliation{Department of Physics, Zhejiang University, Hangzhou 310027, China}

\author{Th. Brückel}

\affiliation{J\"{u}lich Centre for Neutron Science JCNS and Peter Grünberg Institut PGI, JARA-FIT, Forschungszentrum J\"{u}lich GmbH, D-52425 J\"{u}lich, Germany}

\affiliation{J\"{u}lich Centre for Neutron Science JCNS at Heinz Maier-Leibnitz Zentrum (MLZ), Forschungszentrum J\"{u}lich GmbH, Lichtenbergstraße 1, D-85747 Garching, Germany}

\begin{abstract}
Using polarized neutron diffraction and x-ray resonant magnetic scattering (XRMS) techniques, multiple phase transitions were revealed in an underdoped, non-superconducting Eu(Fe$_{1-x}$Ir$_{x}$)$_{2}$As$_{2}$ ($\mathit{x}$ = 0.06) single crystal. Compared with the parent compound EuFe$_{2}$As$_{2}$, the tetragonal-to-orthorhombic structural phase transition and the antiferromagnetic order of the Fe$^{2+}$ moments are significantly suppressed to $\mathit{T_{S}}$ = 111 (2) K and $\mathit{T_{N,Fe}}$= 85 (2) K by 6\% Ir doping, respectively. In addition, the Eu$^{2+}$ spins order within the $\mathit{ab}$ plane in the A-type antiferromagnetic structure similar to the parent compound. However, the order temperature is evidently suppressed to $\mathit{T_{N,Eu}}$= 16.0 (5) K by Ir doping.
Most strikingly, the XRMS measurements at the Ir $\mathit{L_{3}}$ edge demonstrates that the Ir 5$\mathit{d}$ states are also magnetically polarized, with the same propagation vector as the magnetic order of Fe. With $\mathit{T_{N,Ir}}$ = 12.0 (5) K, they feature a much lower onset temperature compared with $\mathit{T_{N,Fe}}$. Our observation suggests that the magnetism of the Eu sublattice has a considerable effect on the magnetic nature of the 5$\mathit{d}$ Ir dopant atoms and there exists a possible interplay between the localized Eu$^{2+}$ moments and the conduction $\mathit{d}$-electrons on the FeAs layers.

\end{abstract}


\maketitle

\section{Introduction}

Among various parent compounds of the recently discovered Fe-based superconductors, \cite{Kamihara_08} EuFe$_{2}$As$_{2}$ is a unique member of the ternary ``122'' $\mathit{A}$Fe$_{2}$As$_{2}$ ($\mathit{A}$ = Ba, Sr, Ca, etc) family since it contains two magnetic sublattices. In a purely ionic picture, the $\mathit{A}$ site is occupied by the $\mathit{S}$-state rare-earth Eu$^{2+}$ ion possessing a 4$\mathit{f}$$^{7}$ electronic configuration with an electron spin $\mathit{S}$ = 7/2, corresponding to a theoretical effective magnetic moment of 7.94 $\mathit{\mu_{B}}$.\cite{Marchand_78} With decreasing temperature, EuFe$_{2}$As$_{2}$ undergoes an antiferromagnetic spin-density-wave (SDW) transition in the Fe sublattice concomitant with a tetragonal-to-orthorhombic structural phase transition at 190 K. In addition, the localized Eu$^{2+}$ spins order in an A-type antiferromagnetic (A-AFM) structure (ferromagnetic (FM) layers ordering antiferromagnetically along the $\mathit{c}$ direction) below 19 K.\cite{Herrero-Martin_09,Xiao_09,Jiang_09_NJP} Similar to other parent compounds of the iron pnictides, chemical substitution \cite{Jeevan_08,Ren_09,Jiang_09,Jiao_11,Jiao_13} or application of external pressure \cite{Miclea_09,Terashima_09} can lead to superconductivity in this system. 

Recently, superconductivity was observed in $\mathit{\textrm{5}d}$ transition metal element doped Eu(Fe$_{1-x}$Ir$_{x}$)$_{2}$As$_{2}$ with $\mathit{T_{SC}}$ up to \textasciitilde{} 22 K.\cite{Paramanik_13,Jiao_13} The magnetic ground state of the optimally doped, superconducting Eu(Fe$_{0.88}$Ir$_{0.12}$)$_{2}$As$_{2}$ was determined by our single-crystal neutron diffraction measurement.\cite{Jin_15} Below 17 K, the Eu$^{2+}$spins order ferromagnetically along the crystallographic $\mathit{c}$-direction. Both the structural phase transition and the SDW order of the Fe sublattice were found to be fully suppressed in the optimally doped compound with bulk superconductivity. Similar conclusions were obtained on polycrystalline Eu(Fe$_{0.86}$Ir$_{0.14}$)$_{2}$As$_{2}$ by Anand $\mathit{et}$ $\mathit{al.}$ based on muon spin relaxation ($\mu SR$) and neutron powder diffraction measurements.\cite{Anand_15} However, the phase diagram describing how the magnetic order of the Eu$^{2+}$ spins is tuned by the Ir doping is still not available. In addition, although it is well established that for the iron pnictides the chemical doping suppresses the SDW order of the Fe$^{2+}$ moments in the underdoped region of their phase diagrams, the magnetic properties of the transition metal dopants themselves were not well understood so far. To the best of our knowledge, there exists only a few experimental studies about the magnetic nature of the transition metal dopants in the iron pnictides.\cite{Sefat_09,Dean_12,Kim_13} For instance, 3$\mathit{d}$ Co in BaCo$_{2}$As$_{2}$ does not order magnetically,\cite{Sefat_09} while spin polarization of 4$\mathit{d}$ Ru and 5$\mathit{d}$ Ir dopant atoms was observed by x-ray resonant magnetic scattering measurements in superconducting Ba(Fe$_{0.795}$Ru$_{0.205}$)$_{2}$As$_{2}$ and Ba(Fe$_{0.973}$Ir$_{0.027}$)$_{2}$As$_{2}$, respectively, revealing strong coupling between the magnetism of Fe and the transition metal dopants.\cite{Dean_12,Kim_13} In our case of Eu(Fe$_{1-x}$Ir$_{x}$)$_{2}$As$_{2}$ contaning both Fe and Eu magnetic sublattices, it will be more interesting to investigate the magnetism of the dopant atoms since both Fe and Eu might exert some influence on them.

Here we present the results of our x-ray resonant magnetic scattering (XRMS) and polarized neutron diffraction measurements on an underdoped, non-superconducting Eu(Fe$_{1-x}$Ir$_{x}$)$_{2}$As$_{2}$ ($\mathit{x}$ = 0.06) single crystal, which displays multiple phase transitions. Compared with the parent compound, the tetragonal-to-orthorhombic structural phase transition and the antiferromagnetic order of the Fe$^{2+}$ moments are well seperated, significantly suppressed to $\mathit{T_{S}}$ = 111 (2) K and $\mathit{T_{N,Fe}}$= 85 (2) K by 6\% Ir doping, respectively. In addition, the Eu$^{2+}$ spins order within the $\mathit{ab}$ plane in the A-type antiferromagnetic structure similar to the parent compound, suggesting that the magnetic structure of the Eu$^{2+}$ spins starts to be tuned from AFM to FM within the intermediate Ir doping level between 6\% to 12\%. However, the ordering temperature is evidently suppressed to $\mathit{T_{N,Eu}}$= 16.0 (5) K by 6\% Ir doping. Most strikingly, the XRMS measurements at the Ir $\mathit{L_{3}}$ edge demonstrates that the Ir 5$\mathit{d}$ states are magnetically polarized with the same propagation vector as the magnetic order of Fe. With $\mathit{T_{N,Ir}}$ = 12.0 (5) K, they feature a much lower onset temperature compared with $\mathit{T_{N,Fe}}$. Our observation suggests that the magnetism of the Eu sublattice has a considerable effect on the magnetic nature of the 5$\mathit{d}$ Ir dopant atoms and there exists a possible interplay between the localized Eu$^{2+}$ moments and the conduction $\mathit{d}$-electrons on the FeAs layers.

\section{Experimental Details}

Single crystals of Eu(Fe$_{1-x}$Ir$_{x}$)$_{2}$As$_{2}$ ($\mathit{x}$ = 0.06) were grown from self-flux (Fe, Ir)As. The chemical composition of the crystals was determined by energy dispersive x-ray (EDX) analysis. A 3 mg platelike single crystal with dimensions \textasciitilde{} 2 $\times$ 2 $\times$ 0.2 mm$^{3}$ was selected for both the XRMS and polarized neutron diffraction measurements. The mosaicity of the crystal was less than 0.03°, confirming the high quality of the chosen crystal. For macroscopic characterizations, the same crystal was used. 

The polarized neutron diffraction measurements were carried out on the diffuse neutron spectrometer (DNS) at Heinz Maier-Leibnitz Zentrum (MLZ), Garching (Germany).\cite{Footnote} The wavelength of the incident neutrons is 4.544 \AA. Although the neutron absorption effect of Eu at such long wavelength is quite strong, the thin platelike shape of the chosen crystal together with the large moment size (\textasciitilde{} 7 $\mathit{\mu_{B}})$ of the Eu$^{2+}$ spins make the neutron measurements feasible \cite{Xiao_09,Jin_13,Nandi_14_neutron,Jin_15} The crystal was mounted on an aluminum sample holder with very tiny amount of GE varnish. The {[}0, 1, 0{]} direction of the crystal was aligned perpendicular to the horizontal scattering plane so that the ($\mathit{H}$,
0, $\mathit{L}$) reciprocal plane can be mapped out by rotating the sample. Throughout this paper, the orthorhombic notation (space group $\mathit{F}$$\mathit{m}$$\mathit{m}$$\mathit{m}$) will be used for convenience. The incident neutron spins were polarized approximately parallel to the scattering vector $\mathit{\mathit{\mathit{\mathit{Q}}}}$ ($\mathit{x}$-polarization), since in case of multi-detectors covering a larger $\mathit{Q}$-range it is impossible to have the polarization parallel to all different $\mathit{Q}$ vectors simultaneously. The scattering intensities in the spin-flip (SF) and non-spin-flip (NSF) channels were collected, respectively, to conclude about the magnetic ground state of the Eu$^{2+}$ moments. For $\mathit{x}$-polarization, the scattering cross sections for NSF and SF processes read as 
\begin{equation}
\left(\frac{d\sigma}{d\Omega}\right)_{x}^{NSF}\propto N^{*}N+\frac{1}{3}I_{SI}
\end{equation}

and
\begin{equation}
\left(\frac{d\sigma}{d\Omega}\right)_{x}^{SF}\propto M_{\bot Y}^{*}M_{\bot Y}+M_{\bot Z}^{*}M_{\bot Z}+\frac{2}{3}I_{SI}
\end{equation}
respectively, where $N^{*}N$ denotes the coherent nuclear scattering and $I_{SI}$ denotes the total spin incoherent scattering, whereas $M_{\bot Y}^{*}M_{\bot Y}$ and $M_{\bot Z}^{*}M_{\bot Z}$ are the components of the moment parallel and perpendicular to the scattering plane, respectively. The symbol $\bot$ indicates that the magnetic scattering is only sensitive to the component of the moment perpendicular to $\mathit{Q}$. Therefore, using $\mathit{x}$-polarization, the magnetic and nuclear scattering can be completely seperated into the SF and NSF channel, respectively. 

The XRMS measurements were performed at the Fe $\mathit{K}$ edge and Ir $\mathit{L_{3}}$ edge at beamline I16 at the Diamond Light Source (Oxford, UK).\cite{Collins_10} The incident radiation was linearly polarized parallel to the horizontal scattering plane ($\pi$-polarization) with the beam size of 0.2 mm (horizontal) $\times$ 0.03 mm (vertical). The magnetic reflections were probed in the $\mathit{\pi-\sigma'}$ scattering channel in which the polarization of the diffracted beam is perpendicular to the scattering plane. The crystal was mounted on a Cu sample holder with some silver paint and then mounted in a closed-cycle cryostat on a six-circle kappa diffractometer.

\section{Experimental Results }

\subsection{Macroscopic characterizations}

Figure 1(a) and (b) show the temperature dependences of the molar specific heat and the in-plane resistivity of the Eu(Fe$_{0.94}$Ir$_{0.06}$)$_{2}$As$_{2}$ single crystal measured using a Quantum Design physical property measurement system (PPMS), respectively. Two phase transitions at 16 $\pm$ 0.5 K and 111 $\pm$ 2 K are clearly visible, corresponding to the antiferromagnetic ordering temperature of the Eu$^{2+}$ moments ($\mathit{T_{N,Eu}}$) and the structural phase transition temperature ($\mathit{T_{S}}$), respectively, as determined by our neutron and x-ray measurements presented below. Another transition at 82 $\pm$ 1 K is also visible in the molar specific heat, as shown in the right inset of Fig. 1(a), which is more evident in its first derivative with respect to the temperature ($\mathit{dC_{p}/dT}$, the red solid curve). We denote this transition temperature as $\mathit{T_{N,Fe}}$ since it coincides with the antiferromagnetic transition determined by our XRMS measurement as discussed in Section III. C. However, this transition is difficult to be resolved in the susceptibility measurement shown in Fig. 1(c) due to small moment size of Fe and the dominant effect of the paramagnetic susceptibility of Eu above $\mathit{T_{N,Eu}}$. In contrast to optimally doped composition Eu(Fe$_{0.88}$Ir$_{0.12}$)$_{2}$As$_{2}$, Eu(Fe$_{0.94}$Ir$_{0.06}$)$_{2}$As$_{2}$ lies in the underdoped side of the phase diagram and it does not exhibit zero resistivity with temperature down to 2K. Nevertheless, a hump is observed in its resistivity curve at low temperature, which is associated with the magnetic ordering of the Eu$^{2+}$ spins. Fig. 1(c) shows the magnetic susceptibility of the Eu(Fe$_{0.94}$Ir$_{0.06}$)$_{2}$As$_{2}$ single crystal under an applied field of 0.1 T perpendicular and parallel to the $\mathit{c}$ direction, respectively, measured using a Quantum Design magnetic property measurement system (MPMS). Below $\mathit{T_{N,Eu}}$, $\mathit{\chi_{ab}}$ drops with decreasing temperature while $\mathit{\chi_{c}}$ remains almost constant, suggesting an antiferromagnetic transtion below which the Eu$^{2+}$ spins might align within the $\mathit{ab}$ plane. Interestingly, there is another kink around $\mathit{T^{*}}$= 11 K in $\mathit{\chi_{ab}}$, whose origin remains unclear so far. Zapf $\mathit{et}$ $\mathit{al.}$\cite{Zapf_13} proposed a spin-glass-transition scenario as a possible explanation for such a similar kink in the in-plane susceptibility of EuFe(As$_{1-x}$P$_{x}$)$_{2}$ single crystals, in which the in-plane component of the Eu$^{2+}$ moments undergo a glassy freezing at lower temperatures compared with $\mathit{T_{N,Eu}}$. 

\begin{figure}
\begin{center}\includegraphics{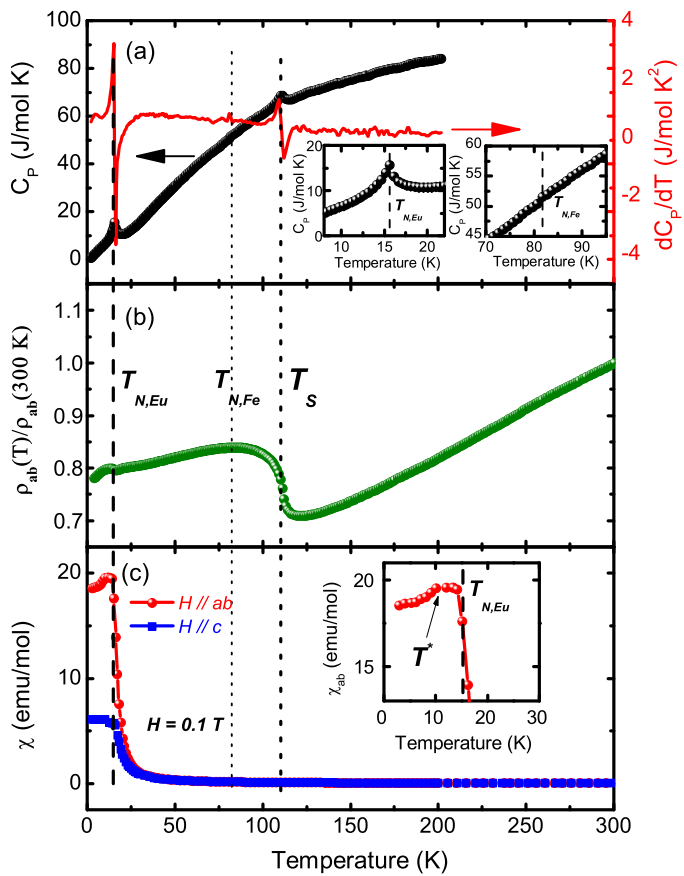}

\caption{(a) The temperature dependence of the molar specific heat, (b) the normalized in-plane resistivity ($\rho_{ab})$, and (c) the molar magnetic susceptibility ($\mathit{\chi})$ of the Eu(Fe$_{0.94}$Ir$_{0.06}$)$_{2}$As$_{2}$ single crystal. The magnetic susceptibility was measured in zero-field-cooling (ZFC) process with an applied field of 0.1 T perpendicular and parallel to the $\mathit{c}$ direction, respectively. The dashed and dotted curves mark the antiferromagnetic ordering temperature of the Eu$^{2+}$ moments ($\mathit{T_{N,Eu}}$) and the structural phase transition temperature ($\mathit{T_{S}}$), respectively. There is another kink around $\mathit{T^{*}}$= 11 K in $\mathit{\chi_{ab}}$, whose origin remains unclear so far.}
\end{center}
\end{figure}

\subsection{Polarized neutron diffraction }

Figure 2(a - c) show the reciprocal space contour maps measured at $\mathit{T}$ = 3.5 K obtained via polarized neutron diffraction. For the neutron polarization parallel to the scattering vector $\mathbf{Q}$ ($\mathit{x}$-polarization), the magnetic and nuclear scattering
intensities can be completely seperated into the SF and NSF channel, as shown in Fig. 2(a) and (b), respectively. The (0, 0, -1) and (0, 0, -3) reflections appear at the base temperature in the SF channel, indicating that magnetic order of the Eu$^{2+}$ spins in Eu(Fe$_{0.92}$Ir$_{0.08}$)$_{2}$As$_{2}$ is antiferromagnetic, similar to the case of the parent compound. In addition, no magnetic intensity is observed at the $\mathbf{Q}$ = (-2, 0, 0) within the experimental resolution (Fig. 2(c)), excluding the possibility of net ferromagnetic (FM) component of the Eu$^{2+}$ spins along the $\mathit{c}$-axis. Such a scenario was suggested by Zapf. $\mathit{et}$ $\mathit{al.}$ \cite{Zapf_11} as a possible intermediate magnetic state for the doped compositions between the parent compound with the Eu$^{2+}$ moments lying completely in the $\mathit{ab}$ plane and high doping levels with the Eu$^{2+}$ moments ordering ferromagnetically along the $\mathit{c}$ axis. Thus the Eu$^{2+}$ magnetic structure here in the Eu(Fe$_{0.94}$Ir$_{0.06}$)$_{2}$As$_{2}$
single crystal is identical to that in the parent compound and 6\% Ir-substitution for Fe seems not enough to change the magnetic ground state of the Eu sublattice. Considering the FM structure determined previously in Eu(Fe\textsubscript{0.88}Ir\textsubscript{0.12})\textsubscript{2}As\textsubscript{2}via unpolarized neutron diffraction,\cite{Jin_15} the change of the Eu$^{2+}$ magnetic structure in the Eu(Fe$_{1-x}$Ir$_{x}$)$_{2}$As$_{2}$ system from the A-type AFM order to the FM order is expected to occur at intermediate Ir doping level between 6\% and 12\%. The temperature dependence of the (0, 0, -3) magnetic reflection is shown in Fig. 2(d). The antiferromagnetic transition temperature $\mathit{T_{N,Eu}}$ is determined to be 16 $\pm$ 0.5 K, consistent with the values expected from the macroscopic measurements. The order parameter can be fitted to the form ($\mathit{T}$ - $\mathit{T_{N,Eu}}$)$^{2\beta}$ with$\mathit{\beta}$ = 0.32(3), which is not precisely determined in the critical region around the magnetic phase transition so not a real ``critical exponent''. But it is quite close to the critical exponent of the three-dimensional Ising model ($\mathit{\beta}$ = 0.326). The temperature dependence of an allowed Bragg reflection, (0, 0, -2), is also shown as reference. 

\begin{figure}
\begin{center}\includegraphics{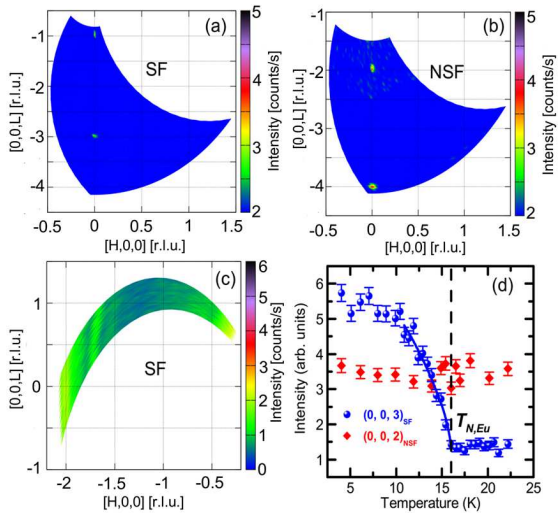}
\caption{(a, b, c) Contour maps in the ($\mathit{H}$, 0, $\mathit{L}$) reciprocal plane for Eu(Fe$_{0.94}$Ir$_{0.06}$)$_{2}$As$_{2}$ at $\mathit{T}$ = 3.5 K obtained via polarized neutron diffraction with neutron polarization parallel to the scattering vector $\mathbf{\mathit{Q}}$ ($\mathit{x}$-polarization). Intensities in the SF channel ((a) and (c)) exclusively correspond to the magnetic reflections, while intensities in the NSF channel (b) solely originate from the nuclear reflections.\cite{Footnote2} (d) The temperature dependencies of the magnetic and nuclear reflections. The solid line in (d) represents a fit of the antiferromagnetic order parameter close to the transition using a power law.}
\end{center}
\end{figure}

\subsection{X-ray resonant magnetic scattering }

The Eu(Fe$_{0.94}$Ir$_{0.06}$)$_{2}$As$_{2}$ single crystal was also measured using synchrotron x-ray diffraction to study its structural properties and the magnetism of the Fe sublattice. Figure 3(a) displays the ($\mathit{H}$, 0, 8) scans through the (4, 0, 8) Bragg reflection at several temperatures. A single (2, 2, 8)$_{T}$ peak in the tetragonal ($\mathit{T}$) phase at 113 K splits into two distinct peaks {[}(4, 0, 8)$_{O}$ and (0, 4, 8)$_{O}$ {]} in the orthorhombic ($\mathit{O}$) phase below $\mathit{T_{S}}$ = 111 K, due to the structural phase transition from $\mathit{I}$$\mathit{4/m}$$\mathit{m}$$\mathit{m}$ to $\mathit{F}$$\mathit{m}$$\mathit{m}$$\mathit{m}$ space group. As plotted in Fig. 3(b), the orthorhmbic distortion $\delta$ = $\mathit{(a}-b)$ / $\mathit{(a}+b)$ increases monotonically below $\mathit{T_{S}}$. $\mathit{T_{S}}$ determined here is well consistent with the anomaly shown in both the specific heat and the resistivity measurements. The abrupt occurrence of the orthorhombic splitting below $\mathit{T_{S}}$ indicates the first-order nature of the structural phase transition.

\begin{figure}
\centering{} \includegraphics{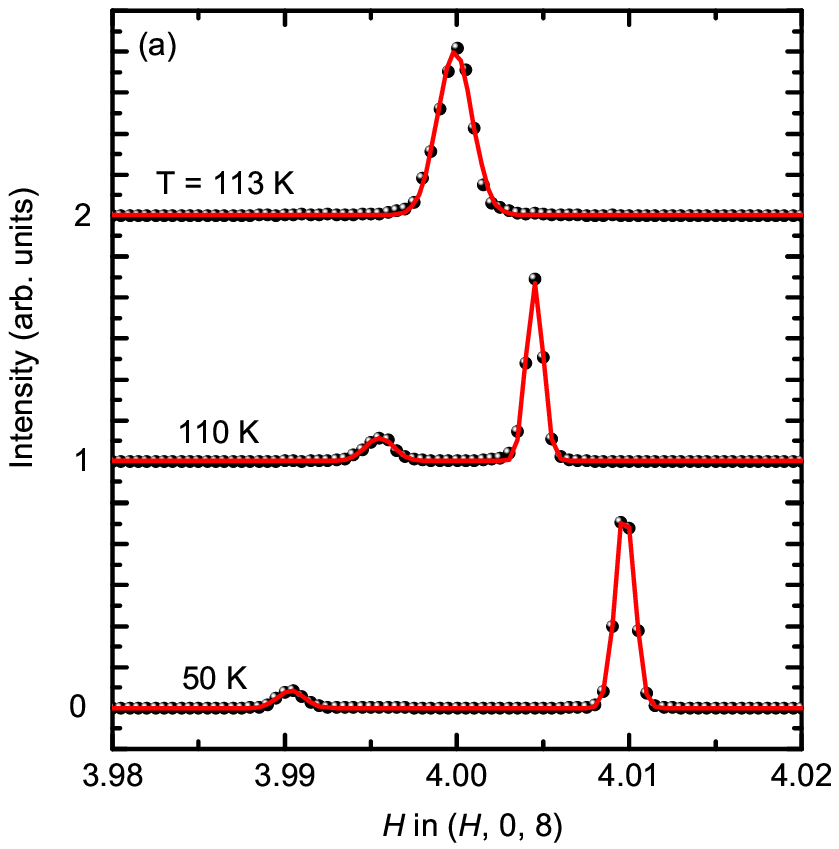}

\includegraphics{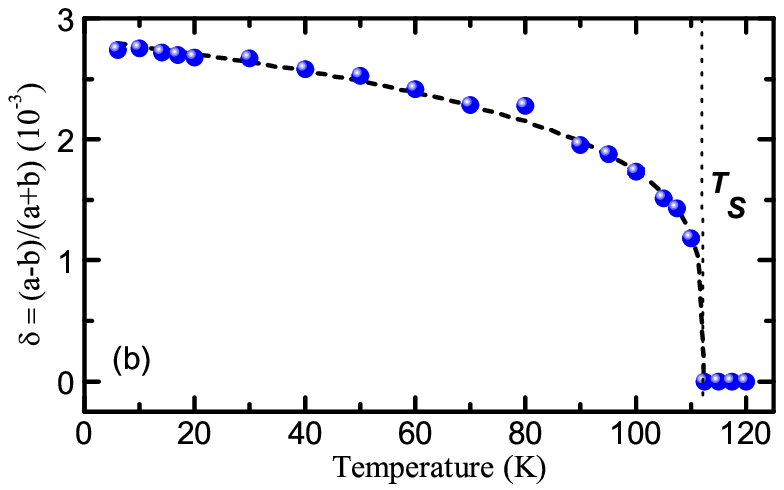}

\caption{(a) $\mathit{H}$-scans through the (4, 0, 8) Bragg reflection at several temperatures, showing the peak splitting due to the tetragonal-to-orthorhombic structural phase transition. (b) The temperature dependence of the orthorhombic distortion parameter$\delta$ = $\mathit{(a}-b)$ / $\mathit{(a}+b)$. The dotted vertical line marks the structural phase transition temperature $\mathit{T_{S}}$ and the dashed curve is a guide to the eyes.}
\end{figure}

As shown in Figure 4(a), at $\mathit{T}$ = 7 K, a magnetic reflection from the Fe$^{2+}$ moments was clearly observed in the $\mathit{\pi-\sigma'}$ scattering channel when the energy of the x-ray was tuned through the Fe $\mathit{K}$ edge (E = 7.111 keV) at $\mathbf{Q}$ = (1, 0, 9). This signal arises from the spin-density-wave (SDW) type antiferromagnetic order of the Fe$^{2+}$ spins characterized by the propagation vector $\mathbf{k}$ = (1, 0, 1), similar to the one observed in the parent compound EuFe$_{2}$As$_{2}$ by nonresonant magnetic x-ray scattering.\cite{Herrero-Martin_09} The peak is displaced from $\mathit{H}$ = 1 due to the orthorhombic distortion below $\mathit{T_{S}}$. However, this peak disappears at 100 K, as shown by the rocking-curve scans in Fig. 4(b). The temperature dependence of the integrated intensity of the (1, 0, 9) magnetic peak normalized to the (2, 0, 10) charge reflection is plotted in Fig. 4(c). The antiferromagnetic transition temperature of the Fe$^{2+}$ moments can be determined as $\mathit{T_{N,Fe}}$ = 85 $\pm$ 2 K. To confirm the resonant magnetic behavior of the peak, energy scans at the Fe $\mathit{K}$ edge were performed. Fig. 4(d) shows the background-subtracted energy scan through the (1, 0, 9) reflection at T = 7 K. The energy spectrum here is very similar to that observed in previous XRMS measurements at the Fe $\mathit{K}$ edge for SmFeAsO and BaFe$_{2}$As$_{2}$.\cite{Nandi_11,Kim_10} This includes a sharp resonant feature close to the absorption threshold, whose energy is consistent with the pre-edge hump observed in the fluorescence spectrum from the sample (Fig. 4(e)), and broad resonant features extending to energies more than 20 eV above the absorption edge. Note that the antiferromagnetic transition of the Fe$^{2+}$ moments occurs at a much lower temperature compared with the structural phase transition ($\mathit{T_{S}}$ = 111 $\pm$ 2 K), similar to the observations in Co-doped EuFe$_{2}$As$_{2}$ and BaFe$_{2}$As$_{2}$ in which the two transitions become well seperated with doping.\cite{Jin_13,Pratt_09,Christianson_09} The existence of both, the structural phase transition and the Fe-SDW order in underdoped, non-superconducting Eu(Fe$_{0.94}$Ir$_{0.06}$)$_{2}$As$_{2}$, is in stark contrast to the case of superconducting Eu(Fe$_{0.88}$Ir$_{0.12}$)$_{2}$As$_{2}$, in which both transitions are completely suppressed in favor of bulk superconductivity.\cite{Jin_15} \begin{figure}
\centering{}\includegraphics{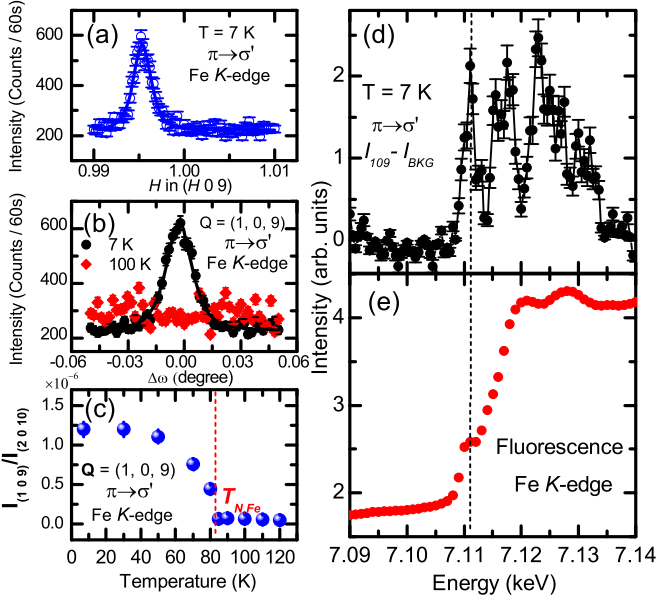}

\caption{(a) $\mathit{H}$ scan through the (1, 0, 9) magnetic reflection of the Fe$^{2+}$ moments in Eu(Fe$_{0.94}$Ir$_{0.06}$)$_{2}$As$_{2}$ at the Fe $\mathit{K}$ edge at $\mathit{T}$ = 7 K. The solid curve represents the fit using a Lorentzian-squared line shape. (b) Rocking-curve scans at $\mathbf{Q}$ = (1, 0, 9) at 7 K and 100 K, respectively, at the Fe $\mathit{K}$ edge. (c) The temperature dependence of the integrated intensity of the (1, 0, 9) magnetic peak normalized to the (2, 0, 10) charge reflection at the Fe $\mathit{K}$ edge. The
short dashed line marks $\mathit{T_{N,Fe}}$, the antiferromagnetic transition temperature of the Fe$^{2+}$ moments. (d) Background-subtracted energy scan through the (1, 0, 9) reflection at T = 7 K. The background was taken at $\mathbf{Q}$ = (0.9, 0, 9). (e) Energy scan of the fluorescence yield at the Fe $\mathit{K}$ edge. The dashed line in (d) and (e) marks the energy at which the measurements in (a), (b) and (c) were done. }
\end{figure}

Furthermore, in order to probe the Ir 5$\mathit{d}$ dopant states, the energy of the incident x-ray was tuned to the Ir $\mathit{L_{3}}$ edge (E = 11.22 keV). Interestingly, at $\mathit{T}$ = 7 K, a clear peak is also present in the $\mathit{\pi-\sigma'}$ scattering channel
at $\mathbf{Q}$ = (1, 0, 9), as shown in Figure 5(a). The propagation vector at which the scattering is observed is identical to that of the antiferromagnetic order of the Fe$^{2+}$ moments, suggesting that this peak most likely arises from the magnetic order of the Ir dopant atoms. Surprisingly, this peak disappears at 13 K, as shown by the rocking-curve scans in Fig. 5(b). The temperature dependence of the integrated intensity of this peak normalized to the (2, 0, 10) charge reflection is plotted in Fig. 5(c). Strikingly, the peak gets suppressed quickly upon heating and disapears completely above $\mathit{T_{N,Ir}}$ = 12 $\pm$ 0.5 K, a temperature much lower than $\mathit{T_{N,Fe}}$= 85 $\pm$ 2 K. This is very surprising and will be discussed below. The background-subtracted energy scan shown in Fig. 5(d) confirms the resonant behavior of the (1, 0, 9) peak at the Ir $\mathit{L_{3}}$ edge. Below the absorption edge (determined by the white line of the fluorescence shown in Fig. 5(e)), there is almost no intensity. When the incident x-ray energy is tuned through the edge, the intensity increases sharply. Above the edge, the intensity shows a tendency to slowly drop. The non-Lorentzian line shape observed here is very similar to that observed at the Ir $\mathit{L_{3}}$ edge using XRMS for superconducting Ba(Fe$_{1-x}$Ir$_{x}$)$_{2}$As$_{2}$. It was attributed to the interference between Ir resonant scattering and Fe nonresonant mangetic scattering.\cite{Dean_12} By fitting to the peak profiles in Fig. 4(a) and 5(a) using  Lorentzian-squared line shape, the full-width at half-maximum (FWHM) of the (1, 0, 9) peak at the Fe $\mathit{K}$ edge and Ir $\mathit{L_{3}}$ edge are revealed to be quite similar, close to 0.002 r. l. u. (reciprocal lattice units). The correlation length ($\zeta$) along the $\mathit{a}$ axis for both Ir and Fe magnetic order can be roughly estimated as: $\zeta$$_{Fe}$ $\approx$ $\zeta$$_{Ir}$ = 2/ (a$^{*}$$\times$FWHM) = 885 \AA = 160 unit cells. Therefore, the polarization of the Ir atoms is quite well correlated, with a similar correlation length as that of the Fe spins, suggesting that the Ir dopant atoms are uniformly distributed over a large length scale. 

\begin{figure}
\centering{}\includegraphics{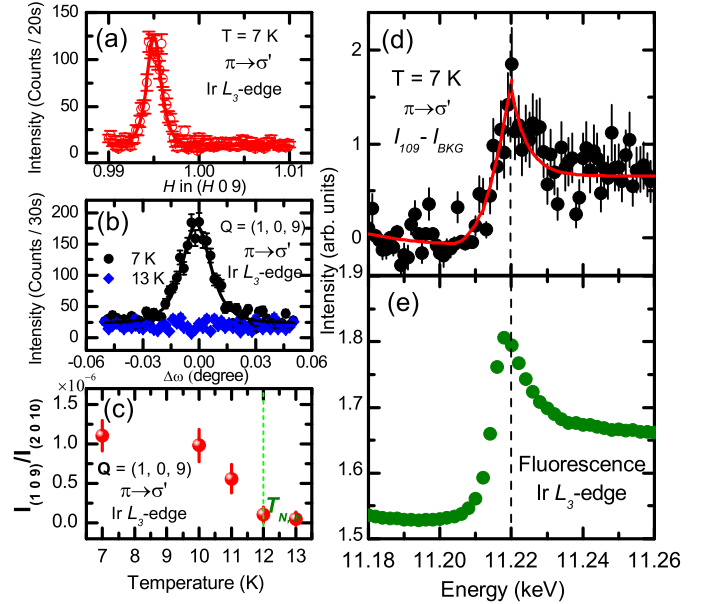}

\caption{(a) $\mathit{H}$ scan through the XRMS peak at the Ir $\mathit{L_{3}}$ edge in Eu(Fe$_{0.94}$Ir$_{0.06}$)$_{2}$As$_{2}$ at $\mathit{T}$ = 7 K. The solid curve represents the fit using a Lorentzian-squared line shape. (b) Rocking-curve scans at $\mathbf{Q}$ = (1, 0, 9) at 7 K and 13 K, respectively, at the Ir $\mathit{L_{3}}$ edge. (c) The temperature dependence of the integrated intensity of the (1, 0, 9) peak normalized to the (2, 0, 10) charge reflection. The short dashed line marks $\mathit{T_{N,Ir}}$, the antiferromagnetic transition temperature of Ir. (d) Background-subtracted energy scan through the (1, 0, 9) reflection at T = 7 K. The background was taken at $\mathbf{Q}$ = (0.9, 0, 9). The solid line is a guide to the eye. (e) Energy scan of the fluorescence yield at the Ir $\mathit{L_{3}}$ edge. The dashed line in (d) and (e) marks the energy at which the measurements in (a), (b) and (c) were done. }
\end{figure}

\section{Discussion And Conclusion}

First of all, the origin of the resonance at the Ir $\mathit{L_{3}}$ edge in in Eu(Fe$_{0.94}$Ir$_{0.06}$)$_{2}$As$_{2}$ needs to be understood. Note that the resonant x-ray scattering signal in the $\mathit{\pi-\sigma'}$ channel can originate from either the spin or orbital ordering and the spin and orbital degrees of freedom are strongly coupled for 5$\mathit{d}$ Ir. Therefore, possible orbital contribution from Ir to the observed (1, 0, 9) peak can't be excluded. In addition, it can also originate from some local structural distortion or preferred neighborhood around the Ir atoms, which makes the scattering amplitude of Ir anisotropic and leads to the resonant scattering at the Ir $\mathit{L_{3}}$ edge. Nevertheless, we speculate that it most likely results from the magnetic polarization of the Ir 5$\mathit{d}$ states, due to two reasons. First, the resonant scattering set in below a relatively low temperature (\textasciitilde{} 12 K),\cite{Footnote4} which is coincidentally close and comparable to the A-type AFM transition temperature of Eu, $\mathit{T_{N,Eu}}$=16 $\pm$ 0.5 K. The discrepancy between $\mathit{T_{N,Ir}}$ and $\mathit{T_{N,Eu}}$ is likely due to the difference of the sample-heating effect between synchrotron x-rays and neutrons. It is noteworthy to point out that $\mathit{T_{N,Ir}}$ is determined from the synchrotron measurement and there is a strong sample-heating effect from the incident x-ray beam if no attenuators are applied in order to probe very weak effect,\cite{Footnote3} while$\mathit{T_{N,Eu}}$ is determined from the neutron diffraction measurement and the sample-heating effect is negligible. Note that we only observed a single anomaly around 16 K at low temperature in the heat capacity measurement shown in Fig.1(a) and no hint for another transition around 12 K can be discernible. Second, the resonant scattering occurs at the same wave vector as is typically observed for the magnetic order in the 122 iron pnictides, with the propagation vector $\mathbf{k}$ = (1, 0, 1).\cite{Su_09,Xiao_09,Goldman_08} 

For superconducting Ba(Fe$_{1-x}$Ir$_{x}$)$_{2}$As$_{2}$, it was proposed that the Ir 5$\mathit{d}$ states may be polarized by either the local field from the Fe neighbors or by other indirect interactions between the Ir and Fe states and it was found by XRMS that the Ir
magnetic order persists up to the Néel transition of the majority Fe spins.\cite{Dean_12} In Eu(Fe$_{0.94}$Ir$_{0.06}$)$_{2}$As$_{2}$, however, clear polarization of Ir occurs at a much lower temperature $\mathit{T_{N,Ir}}$ = 12  $\pm$ 0.5 K compared with that of the majority Fe spins, $\mathit{T_{N,Fe}}$ = 85  $\pm$ 2 K. Comparison between Ba(Fe$_{1-x}$Ir$_{x}$)$_{2}$As$_{2}$ and Eu(Fe$_{1-x}$Ir$_{x}$)$_{2}$As$_{2}$ suggests a considerable effect of the $\mathit{A}$-site ion on the magnetic nature of the 5$\mathit{d}$ Ir dopant atoms. Since the XRMS at the Ir $\mathit{L_{3}}$ edge corresponds to the excitation of the 2 $p_{3/2}$ core electrons into the 5$\mathit{d}$ valence band of Ir, which is hybridized with the 3$\mathit{d}$ valence band of Fe through the chemical doping process as revealed by our previous electronic structure calculation performed on Eu(Fe$_{1-x}$Ir$_{x}$)$_{2}$As$_{2}$,\cite{Jin_15} the polarization of the 5$\mathit{d}$ states of Ir implies possible polarization of the Fe 3$\mathit{d}$ band below $\mathit{T_{N,Ir}}$ ($\thickapprox\mathit{T_{N,Eu}}$) induced by the magnetic order of Eu. A similar effect was revealed in a previous study on EuFe$_{2}$(As$_{0.73}$P$_{0.27}$)$_{2}$ using magnetic Compton scattering (MCS) measurements, where the magnetism of Fe is enhanced when the Eu magnetic order sets in.\cite{Ahmed_10} The interplay between the localized Eu$^{2+}$ moments and the conduction $\mathit{d}$-electrons on the FeAs layers were also observed before in the Co-doped EuFe$_{2}$As$_{2}$ based on nuclear magnetic resonance (NMR) and Mössbauer spectroscopy measurements.\cite{Guguchia_11,Blachowski_11} However, in the parent compound, the coupling between the Eu and Fe sublattices was found to be negligible according to previous neutron and nonresonant x-ray magnetic scattering measurements.\cite{Xiao_09,Herrero-Martin_09} This is not contradictory since it was found that the interplay between 3 $d$ and 4 $f$ electrons is tunable by chemical doping in the iron pnictides CeFe$_{1-x}$Co$_{x}$AsO and GdFe$_{1-x}$Co$_{x}$AsO.\cite{Shang_13} In addition, compared with XRMS at the Fe $\mathit{K}$ edge (1 $s$ $\rightarrow$ 4 $p$), XRMS at the Ir $\mathit{L_{3}}$ edge is more sensitive to the change of the $d$-band electrons and thus can probe their interaction with the Eu 4 $f$ electrons more effectively. Therefore we believe that we have observed the evidence for possible interplay between the Eu and Fe sublattices upon 6\% Ir doping.

In addition, doping more Ir into the Fe site is able to tune the magnetic ground state of the localized Eu$^{2+}$ spins, from the A-type AFM order in Eu(Fe\textsubscript{0.94}Ir\textsubscript{0.06})\textsubscript{2}As\textsubscript{2} to the FM order in Eu(Fe\textsubscript{0.88}Ir\textsubscript{0.12})\textsubscript{2}As\textsubscript{2},\cite{Jin_15} via the change of the indirect Ruderman-Kittel-Kasuya-Yosida (RKKY) interaction mediated by the conduction $\mathit{d}$-electrons on the FeAs layers.\cite{Ruderman_54,Kasuya_56,Yosida_57} 

In summary, we have performed the polarized neutron diffraction and x-ray resonant magnetic scattering (XRMS) measurements on a underdoped, non-superconducting Eu(Fe$_{1-x}$Ir$_{x}$)$_{2}$As$_{2}$ ($\mathit{x}$ = 0.06) single crystal and found multiple phase transitions. The tetragonal-to-orthorhombic structural phase transition and the antiferromagnetic order of the Fe$^{2+}$ moments are well seperated, significantly suppressed to $\mathit{T_{S}}$ = 111 (2) and $\mathit{T_{N,Fe}}$= 85 (2) K by 6\% Ir doping, respectively, compared with the parent compound. In addition, the Eu$^{2+}$ spins order within the $\mathit{ab}$ plane in the A-type antiferromagnetic structure similar to that in the parent compound. However, the order temperature is evidently suppressed to $\mathit{T_{N,Eu}}$= 16.0 (5) K by 6\% Ir doping. Most strikingly, the XRMS measurements at the Ir $\mathit{L_{3}}$ edge demonstrates that the Ir 5$\mathit{d}$ states are magnetically polarized with the same propagation vector as the magnetic order of Fe. With $\mathit{T_{N,Ir}}$ = 12.0 (5) K, they feature a much lower onset temperature compared with $\mathit{T_{N,Fe}}$. Our observation suggests that the magnetism of the Eu sublattice has considerable effect on the magnetic nature of the 5$\mathit{d}$ Ir dopant atoms and there exists a possible interplay between the localized Eu$^{2+}$ moments and the conduction $\mathit{d}$-electrons on the FeAs layers.

\begin{acknowledgments}
This work is based on experiments performed at the DNS instrument operated by Jülich Centre for Neutron Science (JCNS) at the Heinz Maier-Leibnitz Zentrum (MLZ), Garching, Germany and at the I16 beamline at the Diamond Light Source, Didcot, UK. W. T. J. acknowledges B. Schmitz and S. Mayr for the technical assistance. G. H. C. acknowleges the support from NSFC (No.11474252). 
\end{acknowledgments}


\begin{thebibliography}{10}
\bibitem{Footnote2}
Note that the magnetic reflection at (1, 0, -3) and (1, 0, -1) from Fe and Ir can not be observed in the SF channel due to its weakness and strong absorption effect of Eu associated with cold neutrons at DNS.
\bibitem{Footnote4}
Even though the value of $\mathit{T_{N,Ir}}$ is close to that of $\mathit{T^{*}}$, we do not think the magnetic polarization of the Ir dopant atoms is responsible for such an evident kink at $\mathit{T^{*}}$ in the susceptibility data shown in Fig. 1(c). The reason is that the magnetic polarization of the Ir dopant atoms is a weak effect as evidenced in our XRMS measurement and the Ir moment is expected to be much smaller (less than 1 $\mathit{\mu_{B}}$ as revealed in other iridates) than that of Eu (close to 7 $\mathit{\mu_{B}}$). Note that even the Néel transition of the majority Fe spins can not be resolved in Fig. 1(c).  As mentioned in Sec. III.A, the origin of the kink at $\mathit{T^{*}}$ remains unclear so far. It might correspond to a short-ranged glassy freezing of the Eu$^{2+}$ moments as proposed by Zapf. $\mathit{et}$ $\mathit{al.}$ (ref. 23), which is then difficult to be identified by diffraction experiments.
\bibitem{Footnote3}
When measuring the (1, 0, 9) resonant peak at the Ir $\mathit{L_{3}}$ edge, we did not apply any attenuation for the synchrotron beam since the resonant signal was quite weak. It is therefore expected that the actual sample temperature might be higher than the measured value due to the non-instantaneous thermal conduction, so that $\mathit{T_{N,Ir}}$ might be somewhat underestimated by our XRMS measurement. In our case, the strength of such beam-heating effect can be roughly estimated to produce a temperature difference of $\delta$T $\approx$ 5.7 K according to the heating power of the incident beam and the thermal conductivity of the sample by ignoring the influence of thermal convection and thermal radiation. Therefore, a difference of 4 K between $\mathit{T_{N,Ir}}$ and $\mathit{T_{N,Eu}}$ is possible at low temperature considering the inaccuracy of our approximation.
\end{thebibliography}
\end{document}